%% file: KazukiKojima.tex
\documentclass{moriond}

\def\PRL{{\em Phys. Rev. Lett.}}
\def\PRD{{\em Phys. Rev.} D}

\def\be{\begin{equation}}
\def\ee{\end{equation}}
\def\bea{\begin{eqnarray}}
\def\eea{\end{eqnarray}}

\usepackage{subfiles}
\usepackage{lipsum}
\graphicspath{{Figures/}{../Figures/}}

\newcommand{\Photo}{}

\begin{document}
\vspace*{4cm}
\title{Belle~II results related to $b \to c$ anomalies}

\author{K. Kojima (on behalf of the Belle II collaboration)}

\address{Department of Physics, Graduate School of Science, Nagoya University, \\
Furo-cho, Nagoya, Aichi, Japan }

\maketitle\abstracts{
\input{Sections/0-Abstract}
}

\section{Introduction}

\subfile{Sections/1-Introduction}

\section{Measurement of $R(X_{e/\mu})$}

\subfile{Sections/2-RXemu}

\section{Measurement of angular asymmetry $\Delta\mathcal{A}_{x}(w)$}

\subfile{Sections/3-AFB}

\section{Conclusion}

\subfile{Sections/4-Conclusion}

\section*{Acknowledgments}

\subfile{Sections/A-Acknowledgements}

\section*{References}

\subfile{Sections/B-References}
\end{document}

%% file: Sections/0-Abstract.tex
We report two new measurements for tests of lepton flavor universality between electrons and muons using semileptonic $B$ decays with a data set of $189~\mathrm{fb}^{-1}$ collected at the Belle II experiment between 2019 and 2021. Firstly, we find a ratio of inclusive branching fractions at $R(X_{e/\mu}) \equiv \mathcal{B}(\overline{B} \rightarrow Xe^{-}\overline{\nu}_{e})/\mathcal{B}(\overline{B} \rightarrow X\mu^{-}\overline{\nu}_{\mu}) = 1.007 \pm 0.009 (\mathrm{stat.}) \pm 0.019 (\mathrm{syst.})$. This inclusive measurement leads to the most precise universality test based on branching fractions. The measured $R(X_{e/\mu})$ is consistent with the Standard-Model prediction. Secondly, we measure a comprehensive set of five angular asymmetries of $\overline{B}^{0} \rightarrow {D^{*}}^{+}\ell^{-}\overline{\nu}_{\ell}$ decays and obtain agreements with the Standard Model for the measured asymmetries and their differences at a $p$-value above $13\%$. From both tests, no evidence of lepton universality violation is found.

%% file: Sections/1-Introduction.tex
The Standard Model (SM) postulates the lepton flavor universality (LFU) that a gauge boson couples equally to all three generations of leptons. Semileptonic $\overline{b}\rightarrow c\ell^{-}\overline{\nu}_{\ell}$ transition mediated by a $W$ boson. The branching fractions are described commonly for all lepton flavors and only depend on charged lepton masses in the SM. The universality has been supported by experimental results in decays of gauge bosons, light mesons, and leptons~\cite{lfu:2014}, \cite{atlaswlfu:2021}. However, recent observations in $\bar{B} \rightarrow {D^{(*)}}\ell\bar{\nu}_{\ell}$ decays indicate a deviation from the SM by more than $3\sigma$~\cite{hflav:2023}. This discrepancy raises suspicions that the LFU could be violated due to processes involving New Physics, such as leptoquarks~\cite{rdnp1:2018}, \cite{rdnp2:2021}. In addition, it motivates to test the LFU between electron and muons since some of New Physics effects can contribute to interactions with light leptons.

To address this, we conduct the LFU test between electrons and muons at the Belle II experiment. The $B$-factory provides favorable conditions for studying semileptonic $B$ decays that contain missing neutrinos in the final states, thanks to low multiplicity and well-known initial kinematics of $e^{+}e^{-}$ collisions. We perform two new measurements: one using a ratio of inclusive branching fractions and another using differences of angular asymmetries. These measurements offer theoretically and experimentally clean probes by major cancellation of uncertainties.

We use an electron-positron collision data of $189~\mathrm{fb}^{-1}$ collected with the Belle II detector between 2019 and 2021 at a center-of-mass energy of $10.58~\mathrm{GeV}$, corresponding to the $\Upsilon(4S)$ resonance. The data contains $(198.0 \pm 3.0) \times 10^{6}~B\overline{B}$ pairs. We also analyze an off-resonance data of $18~\mathrm{fb}^{-1}$ at an energy 60 MeV below the $\Upsilon(4S)$ resonance to evaluate background contributions from $e^{+}e^{-} \rightarrow q\overline{q}$~($q=u,d,s,c$) processes.
The Belle II detector consists of vertex and tracking detectors, particle identification detectors, an electromagnetic calorimeter, a $1.5~\mathrm{T}$ solenoid magnet, and a $K_{L}^{0}$ and muon detector from the innermost to the outermost layers~\cite{b2tdr:2010}.
Monte Carlo (MC) simulation, where reflects the detector responses, is generated to optimize selections and estimate signal and background distributions.
The events from both the data and the simulation are reconstructed with the Belle II Software Framework, \texttt{basf2}~\cite{basf2:2019}.

%% file: Sections/2-RXemu.tex
We measure the ratio of inclusive branching fractions of semileptonic $B$ decays defined by $R(X_{e/\mu}) \equiv \mathcal{B}(\overline{B} \rightarrow Xe^{-}\overline{\nu}_{e}) / \mathcal{B}(\overline{B} \rightarrow X\mu^{-}\overline{\nu}_{\mu})$, where $X$ is $B$-daughter hadrons reconstructed inclusively.
One of the $B$ mesons, referred to as $B_{\mathrm{tag}}$, from a $\Upsilon(4S)$ decay is fully reconstructed through hadronic $B$ decay channels using the Full Event Interpretation (FEI) algorithm~\cite{fei:2019}. The FEI assigns a confidence score between 0 and 1 to each $B_{\mathrm{tag}}$ candidate and we select only a $B_{\mathrm{tag}}$ with the highest score in an event. From the remaining tracks, we find a lepton candidate with a momentum in the signal $B$ rest frame, $p_{\ell}^{B}$, above $1.3~\mathrm{GeV}/c$. The lepton flavor classification is performed using a multiclass boosted decision tree (BDT) for electrons and a likelihood ratio for muons. Other tracks and clusters not used for the $B_{\mathrm{tag}}$ and lepton reconstruction are attributed to the inclusive hadron system of signal $B$ daughters. Two $B$ mesons with opposite and same flavors are combined for signal $B$ and background-enriched control channels, respectively. The events with same $B$ flavors allow for the inclusion of $B\overline{B}$ backgrounds from events with a misidentified lepton candidate or with a correct lepton coming from a secondary decay of a charmed meson, as well as correct signal candidates with $B^{0}$-$\overline{B}^{0}$ mixing. This control channels help to constrain background yields in the signal channels through a simultaneous fit.

The background candidates from $e^{+}e^{-} \rightarrow q\overline{q}$~($q=u,d,s,c$) events are suppressed with a dedicated BDT employing 21 event-topology variables. The BDT rejects $55\%$ of $q\overline{q}$ backgrounds while retaining $97\%$ of $B\overline{B}$ candidates. The $q\overline{q}$ background distributions are described using the off-resonance data, which is scaled according to the luminosity and the $q\overline{q}$ cross-section.

\begin{figure}
    \centering
    \includegraphics[width=\linewidth]{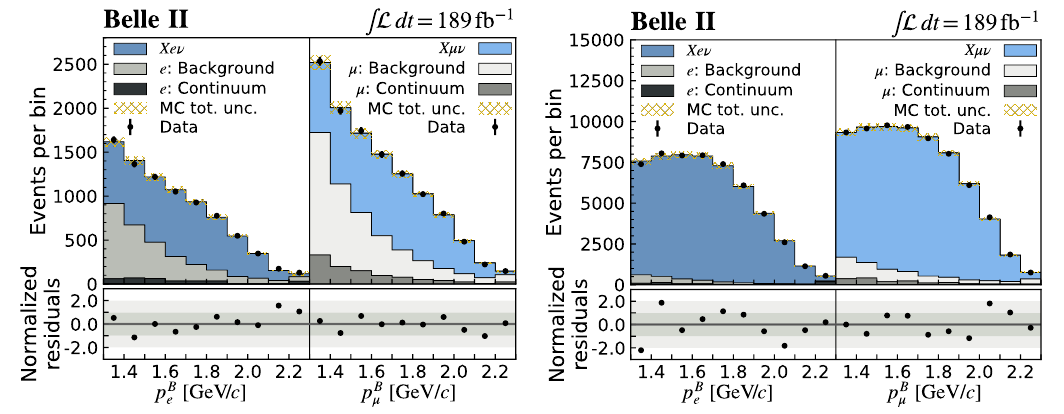}
    \caption{Fit results of lepton momentum spectra in the $B_{\mathrm{sig}}$ rest frame, $p_{\ell}^{B}$~($\ell=e,\mu$), for same-charge control channel (left) and in opposite-charge signal channel (right). Total statistical and systematic uncertainties of the MC samples are shown by yellow hatched areas. Electron and muon modes are illustrated on left and right side of each plot. The last bins of $p_{\ell}^{B}$ include overflow entries.}
    \label{fig:RXemu_postfit}
\end{figure}

The signal yields of both electron and muon modes are extracted with a binned maximum-likelihood simultaneous fit to $p_{\ell}^{B}$ in both the signal and control channels. The signal and $B\overline{B}$ background templates are prepared based on the MC simulation.
The fit yields 
\begin{equation}
    R(X_{e/\mu}) = 1.007 \pm 0.009 (\mathrm{stat.}) \pm 0.019 (\mathrm{syst.})
    \label{eq:RXemu_result},
\end{equation}
and the fit distributions on $p_{\ell}^{B}$ are illustrated in Figure~\ref{fig:RXemu_postfit}.
The most significant systematic uncertainty arises from the lepton identification at $1.9\%$, which is followed by the simulation sample size at $0.9\%$. The uncertainties associated with $\overline{B} \rightarrow X\ell^{-}\overline{\nu}_{\ell}$ branching fractions and $\overline{B} \rightarrow X\ell^{-}\overline{\nu}_{\ell}$ form factors are smaller at $0.2\%$ due to cancellation between electron and muon modes in the $R(X_{e/\mu})$ ratio.
The measured $R(X_{e/\mu})$ value in Eq.~\ref{eq:RXemu_result} agrees with a previous result conducted by the Belle experiment using exclusive $\overline{B} \rightarrow D^{*}\ell^{-}\overline{\nu}_{\ell}$ decays~\cite{bellevub:2022}.  
Furthermore, this result is in agreement with the SM prediction under the LFU~\cite{smrx:2022}. Our $R(X_{e/\mu})$ measurement has provided the most precise result for a test of light-lepton flavor universality based on branching fractions of semileptonic $B$ decays by utilizing inclusive modes for the first time. Further details of our $R(X_{e/\mu})$ measurement can be found in Ref.~\cite{rxemu:2023}.

%% file: Sections/3-AFB.tex
We perform a measurement of angular asymmetries in $B^{0} \rightarrow D^{*}\ell^{-}\overline{\nu}_{\ell}$ decays and their differences between electron and muon modes with a set of five angular observables, $A_{\mathrm{FB}}$, $S_{3}$, $S_{5}$, $S_{7}$, and $S_{9}$. 
The $B^{0} \rightarrow D^{*}\ell^{-}\overline{\nu}_{\ell}$ decays are characterized by a recoil parameter, $w \equiv (m_{{B}^{0}}^{2} + m_{{D^{+}}^{*}}^{2} - (p_{{B}^{0}} - p_{{D^{+}}^{*}})^{2}) / 2m_{B^{0}}m_{{D^{*}}^{+}}$, along with three helicity angles: the angle between the charged lepton in the virtual $W$ frame and the $W$ in the $B^{0}$ frame, $\theta_{\ell}$; the angle between the $D^{0}$ in the ${D^{*}}^{+}$ rest frame and the ${D^{*}}^{+}$ in the $B^{0}$ frame, $\cos{\theta_{V}}$; and the angle between the decay planes formed by the virtual $W$ and the ${D^{*}}^{+}$ in the $B^{0}$ frame, $\chi$.
The angular asymmetries are described by the differential rates of the recoil parameter and helicity angles,
\begin{equation}
    \mathcal{A}_{x}(w) \equiv \left( \frac{\mathrm{d}\Gamma}{\mathrm{d}w} \right)^{-1} \left[ \int_{0}^{1} - \int_{-1}^{0} \right] \mathrm{d}x \frac{\mathrm{d}^{2}\Gamma}{\mathrm{d}w\mathrm{d}x},
    \label{eq:AFB_def}
\end{equation}
where $x = \cos{\theta_{\ell}}$ for $A_{\mathrm{FB}}$, $\cos{2\chi}$ for $S_{3}$, $\cos{\chi}\cos{\theta_{V}}$ for $S_{5}$, $\sin{\chi}\cos{\theta_{V}}$ for $S_{7}$, and $\sin{2\chi}$ for $S_{9}$.
We compare electron and muon modes by calculating the differences, $\Delta\mathcal{A}_{x}(w) = \mathcal{A}_{x}^{\mu}(w) - \mathcal{A}_{x}^{e}(w)$, to test the LFU. $A_{\mathrm{FB}}$, $S_{3}$, and $S_{5}$ exhibit high sensitivity to the LFU violation induced by SM extensions. On the other hand, $S_{7}$ and $S_{9}$ show reduced sensitivity or no sensitivity to the violation, respectively. The latter two observables are taken advantage of as control variables of the analysis method.

One $B$ meson is fully reconstructed with the FEI algorithm, and only one $B$ meson with the highest confidence score is selected in each event.
The remaining tracks and clusters are assigned to the signal $B$ reconstruction. For the $\overline{B}^{0} \rightarrow {D^{*}}^{+}\ell^{-}\overline{\nu}_{\ell}$ decays, we combine a ${D^{*}}^{+}$ candidate with an electron (muon) candidate that is required to have a momentum above $0.4~\mathrm{GeV}$. The lepton flavors are identified by the BDT for electrons and the likelihood ratio for muons. ${D^{*}}^{+}$ candidates are reconstructed through ${D^{*}}^{+} \rightarrow D^{0}\pi^{+}$. $D^{0}$ candidates are reconstructed via eight decay modes of $K^{-}\pi^{+}$, $K^{-}\pi^{+}\pi^{-}\pi^{+}$, $K^{-}\pi^{+}\pi^{0}$, $K^{-}\pi^{+}\pi^{-}\pi^{+}\pi^{0}$, $K_{S}^{0}\pi^{+}\pi^{-}$, $K_{S}^{0}\pi^{+}\pi^{-}\pi^{0}$, $K_{S}^{0}\pi^{0}$, and $K^{+}K^{-}$. The $D^{(*)}$ candidates must satisfy $M_{D^{0}} \in [1.85, 1.88]~\mathrm{GeV}$ ($\Delta M_{{D^{*}}^{+}} \in [0.143, 0.148]~\mathrm{GeV}$), where $\Delta M_{{D^{*}}^{+}} \equiv M_{{D^{*}}^{+}} - M_{D^{0}}$. We require that no additional tracks remain apart from those used for the reconstruction in the event. We select only a single $B\overline{B}$ candidate with $\Delta M_{{D^{*}}^{+}}$ closest to the PDG value~\cite{pdg:2023} for each event.

We extract the signal yields by a binned maximum-likelihood template fit to the $M_{\mathrm{miss}}^{2}$ distributions. 
The extracted yields are corrected for migrations in $M_{\mathrm{miss}}^{2}$ bins, migration of angular and $w$ ranges, efficiencies, and acceptance effects with a migration matrix. To maximize the sensitivity to SM extensions~\cite{afbtheo:2023}, we define three $w$ ranges for integration of angular asymmetries in Eq.~(\ref{eq:AFB_def}) and simultaneously determine all asymmetry variables for both electron and muon modes in different $w$ ranges. The $w$ ranges are set at $[1.000, 1.503]$, $[1.000, 1.275]$, and $[1.275, 1.503]$ for the full ($w_{\mathrm{incl.}}$), low ($w_{\mathrm{low}}$), and high ($w_{\mathrm{high}}$) ranges, respectively.

\begin{figure}
    \centering
    \includegraphics[width=\linewidth]{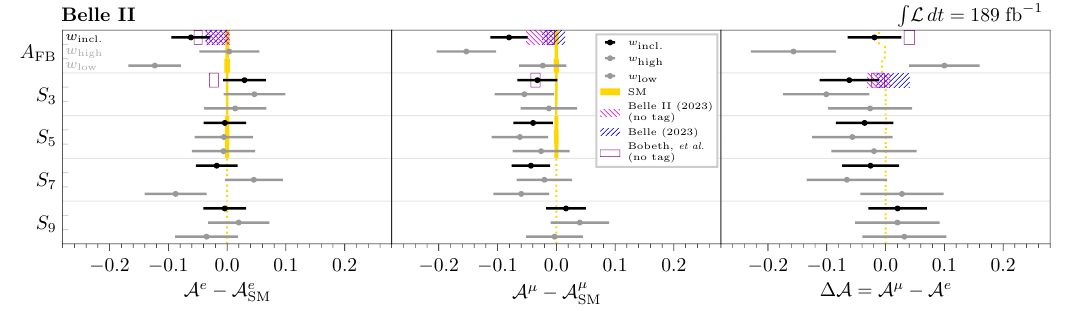}
    \caption{Measured asymmetries for electron (left) and muon (middle) modes and their differences (right). Black and gray points are determined in range of full $w$ range and high or low $w$ ranges, respectively. Yellow solid boxes or dashed lines show the SM prediction.}
    \label{fig:AFB_results}
\end{figure}

We find angular asymmetries for electron and muon modes and their differences as summarized in Figure~\ref{fig:AFB_results}. 
The dominant source of systematic uncertainty is the size of the simulation sample used to estimate the migration matrix, contributing around a quarter to a half of the statistical uncertainties. Other systematic uncertainties, such as lepton identification, remain small up to 0.004.

$\chi^{2}$ tests are performed to evaluate the agreement with the SM predictions for sets of all asymmetries $\mathcal{A}_{x}$, three asymmetry differences of $\Delta A_{\mathrm{FB}}$, $\Delta S_{3}$, and $\Delta S_{5}$, and two asymmetry differences of $\Delta S_{7}$ and $\Delta S_{9}$. These tests returns $\chi^{2}/N_{\mathrm{dof}} = 14.6/10$ $(p=0.14)$, $2.0/3$ $(p = 0.57)$, and $0.6/2$ $(p = 0.76)$ in the $w_{\mathrm{incl.}}$ range, respectively. The tests performed in $w$ subranges, $w_{\mathrm{high}}$ and $w_{\mathrm{low}}$, also yields $p$-values not less than $0.13$. Therefore, there is no evidence of violation of the light-lepton flavor universality observed.

%% file: Sections/4-Conclusion.tex
In the SM, the coupling of gauge bosons is assumed to be common between all lepton flavors as LFU. However, it is recently challenged between tau and light leptons by several experimental results using semileptonic $B$ decays. To test the LFU between electrons and muons, we perform two first measurements utilizing a ratio of inclusive branching fractions, $R(X_{e/\mu})$, and differences of angular asymmetries, $\Delta \mathcal{A}_{x}(w)$, with the $189~\mathrm{fb^{-1}}$ data set collected at the Belle II experiment between 2019 and 2021. We determine $R(X_{e/\mu}) = 1.007 \pm 0.009 (\mathrm{stat.}) \pm 0.019 (\mathrm{syst.})$. The inclusive measurement results in a branching-fraction based LFU test at the world-leading precision and the obtained $R(X_{e/\mu})$ agrees with the SM prediction. In addition, we find the universality for a comprehensive asymmetry set of $A_{\mathrm{FB}}$, $S_{3}$, $S_{5}$, $S_{7}$, and $S_{9}$ with at least a $p$-value of $13\%$. From both tests, no evidence of the LFU violation is found.

%% file: Sections/A-Acknowledgements.tex
This work was supported by Grant-in-Aid for JSPS Fellows Grant Number JP21J15570.